\documentclass[prl,aps,showpacs,twocolumn,preprintnumbers]{revtex4}
\usepackage{amsfonts}
\usepackage{amsmath}
\usepackage{amssymb}
\usepackage{graphicx}%\usepackage{epsfig}
\usepackage{dcolumn}
\begin{document}  
\title{Supersymmetric twisting of carbon nanotubes}
\author{\textsf{V\'{\i}t Jakubsk\'y${}^1$, Mikhail S. Plyushchay${}^{2}$}
 \\
%EndAName
{\small \textit{${}^1$Nuclear Physics Institute, \v Re\v z near Prague, 25068, Czech Republic}}\\
{\small \textit{${}^2$Departamento de F\'{\i}sica, Universidad de
Santiago de Chile, Casilla 307, Santiago 2, Chile}} \\
%{\small \textit{E-mails: v.jakubsky@gmail.com, 
%mikhail.plyushchay@usach.cl} }\\
\date{}
}

\begin{abstract}
We construct exactly solvable models of twisted carbon nanotubes via
supersymmetry, by applying the matrix  Darboux transformation.
We  derive the Green's function for these systems  and compute  the
local density of states. Explicit examples of twisted carbon nanotubes are produced,
where the back-scattering is suppressed and bound states are present. 
We find that  the local density of states decreases in the regions where the bound states are 
localized. Dependence of bound-state energies on the asymptotic twist of the 
nanotubes is determined. We also show that each of the constructed unextended  first order 
matrix systems  possesses a proper nonlinear hidden  supersymmetric structure
with a nontrivial grading operator.
\end{abstract}
\maketitle

\section{Introduction}

Importance of solvable models in physics is enormous. We can acquire
 qualitative understanding of the complicated realistic systems
by analyzing simplified models that grab the essence of a physical reality. 
These models can serve as a test field for approximative
methods,  or can be used as  initial solvable systems in a
perturbative treatment. In this paper, we will focus on the construction 
and analysis of exactly solvable models described by the $(1+1)$-dimensional Dirac equation.

Such systems lie in the overlap of the quantum field theory with the 
condensed matter physics. The one-dimensional Dirac Hamiltonian appears 
in the study of the gap equation of the $1+1$ dimensional version of the 
Nambu-Jona-Lasinio (chiral Gross-Neveu) model \cite{Dunne}, \cite{Feinberg}, 
\cite{Thies}, or in the study of the fractionally charged solitons 
\cite{Jackiw}, \cite{Jackiw2}. It is used in the effective description of 
the non-relativistic fermions:  in \cite{Takayama}, the Hamiltonian describes 
fermions coupled to solitons in the continuum model of a linear molecule of 
polyacetylene. It is employed in the analysis of the quasi-particle bound 
states  associated with the planar solitons in superfluid $^3\mbox{He}$ 
\cite{Ho}. It appears in the description of inhomogeneous superconductors 
\cite{inhomogeneous} and in the analysis of the vortex in the 
extreme type-II superconductors in the mean field approximation \cite{Waxman}.
Last but not least, it is used in the description of carbon nanotubes. In the 
low energy regime, the band structure obtained by tight-binding approach can 
be approximated very well with the use of the one-dimensional Dirac operator 
\cite{Wallace}, \cite{Semenoff}. The stationary equation \cite{Paulimatrix}
\begin{equation}\label{eq1}
 (i\sigma_2\partial_x+\Delta_1(x)\sigma_1)\phi=\lambda\phi
\end{equation}
describes dynamics of the low-energy charge-carriers in single wall carbon 
nanotubes in presence of magnetic field  \cite{Roche}, \cite{KaneMele}.

The Green's function (or its spatial trace called diagonal resolvent or 
Gorkov Green's function) plays an important role in the above mentioned systems. 
It is used in solution of the gap equation \cite{Dunne} or in the extremal 
analysis of the effective action \cite{Feinberg} in quantum field systems. 
It is employed in computation of the free energy of the inhomogeneous 
superconductors \cite{Kos}. It serves in derivation of the local density of states
(LDOS), the quantity that can be measured in carbon nanostructures 
by the spectral tunneling microscopy \cite{STM}, \cite{STM2}. The results 
obtained in this paper will be primarily discussed 
in the latter context.

The carbon nanotubes are cylinders of small radius rolled up from graphene. 
They can be classified as either metallic or semiconducting, in dependence on 
their electronic properties. When no external potential is present, the semi-conducting 
nanotube has a spectral gap which is related to a constant value of the potential, 
$\Delta_1=p_y\neq 0$, where $p_y$ is the value of the canonical momentum in the 
compactified direction. For $\Delta_1=p_y=0$, the nanotube is metallic as it has no 
gap in the spectrum. In this case, an infinitesimally small excitation is sufficient 
to move the electrons from valence to conduction band. The actual value of $p_y$ is 
related to the orientation of the crystal lattice in the nanotube, see e.g., 
\cite{Roche}, \cite{KaneMele}, \cite{nasKlein}.

We suppose that the potential $\Delta_1(x)$ is smooth on the scale of the interatomic 
distance. Otherwise, it would be necessary to work with an extended, $4\times 4$, 
Hamiltonian that would describe mixing of the states between the valleys associated 
with two inequivalent Dirac points \cite{graphene}, \cite{Ando1}.  The matrix degree 
of freedom of $\phi$ in (\ref{eq1}) is the so-called pseudo-spin and is associated 
with the two triangular sublattices that build up the hexagonal structure of the 
graphene crystal; the wave function with either spin-up or -down is identically 
zero on one of the sublattices.

The inhomogeneous magnetic field can appear due to an external source. 
Alternatively, it can emerge as a consequence of mechanical deformations of the 
lattice. Let us make this point clear. Deformation of the lattice is described by 
the vector $\mathbf{d}=(d_x(x,y),d_y(x,y))$ which represents displacement of the 
atoms in the crystal. The associated strain tensor $s_{ij}$ is defined as
\begin{equation}\label{strain}
 s_{xx}=\partial_xd_x,\quad  s_{yy}=\partial_yd_y,\quad s_{xy}=s_{yx}=\frac{\partial_xd_y+\partial_yd_x}{2}.
\end{equation}
The effective Dirac Hamiltonian which describes dynamics of quasi-particles in the 
low-energy regime gets the form 
$\sigma_2(i\partial_x+\Delta_2(x))+\sigma_1(p_y+\Delta_1(x))+\mathbf{1}\Delta_0$, 
where we fixed the Fermi velocity $v_F=1$. 
The gauge fields are related to the strain tensor (\ref{strain}) in this way: 
$\Delta_2(x)= (s_{xx}-s_{yy})$, $\Delta_1(x)= 2s_{xy}$ and $\Delta_0(x)= s_{xx}+s_{yy}$ 
up to multiplicative constants, 
see \cite{KaneMele}, \cite{ando}, \cite{Vozmediano}.  

In this context, the potential $\Delta_1(x)$ in  (\ref{eq1}) can be interpreted as 
the gauge field generated by the twist perpendicular to the axis of the \textit{metallic} 
nanotube. The angle of the twist $\vartheta(x)$  is related to the displacement 
$\mathbf{d}=(0,\int \Delta_1(x)dx)$  by $d_y(x)=r\vartheta(x)$ where  $r$ is a radius 
of the nanotube. In this way, the constant potential $\Delta_1(x)=\beta>0$ can be 
associated with a linear displacement ${\mathbf d}=(0,\beta x)$ that would be generated 
by the constant twist illustrated in Figure 1. It opens a gap in the spectrum of the 
metallic nanotube, however, it does not confine charge carriers. Indeed, constant potential 
can be understood as a mass term in the Hamiltonian describing the free particle.
\newsavebox{\figlinear}        % vytvoření boxu
    \savebox{\figlinear}{          % uložení obrázku do boxu
    \scalebox{1}{
    \includegraphics[scale=.7]{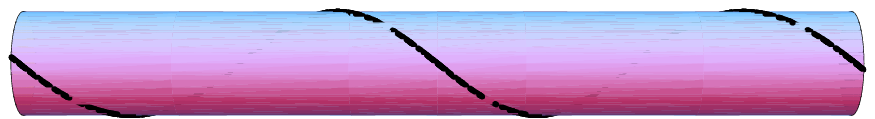}
    }
    }
\begin{figure}\begin{center}\label{figlinear}
\usebox{\figlinear}\caption{The nanotube with the twist corresponding to $ d_y\sim x$. 
In the untwisted nanotube, the black line would be straight (horizontal).}
\end{center}\end{figure}

In general, the electromagnetic field causes nontrivial scattering of the quasi-particles 
and can even cause the appearance of bound states in the system \cite{Hartmann}. It is 
well known that the quasi-particles in metallic nanotubes are not backscattered by 
electrostatic potential. This is understood as a manifestation of the Klein tunneling 
\cite{Klein} and it has been discussed extensively in the literature  
\cite{Ando1}, \cite{KatsnelsonKlein}. It was found recently that the phenomenon can be attributed
to the peculiar supersymmetric structure that relates the
Hamiltonian of the system to that of the free Dirac particle \cite{nasKlein}. 

Here,  we will construct exactly solvable
models described by (\ref{eq1}) where, despite the presence of the effective magnetic 
field, the scattering will be reflectionless and the bound states will be confined 
in the regions where the twist gets altered. In the construction, the techniques known 
in the supersymmetric quantum mechanics will be employed. We will focus to the spectral 
properties and Green's function of the new systems. The latter one will be used for 
computation of the LDOS. We will provide an analytical formula for bound state energies 
in dependence on the twist of the nanotubes.  

The work is organized as follows. In the next section, we briefly review the construction 
of solvable models based on Darboux transformation with focus on the application in the 
context of carbon nanotubes. Then the formulas for Green's function and LDOS of these 
models are provided. We discuss reflectionless systems and present two models of twisted 
carbon nanotubes. The last section is left for the discussion.

\section{Spectral design via Darboux transformations}

We summarize here the main points of the construction of new solvable models which is 
based on the intertwining relations.
This scheme is well known in the context of supersymmetric (SUSY) quantum mechanics 
\cite{JunkerKhare}. There, the intertwined second order Schr\"odinger operators give 
rise to the supersymmetric Hamiltonian while the intertwining operator, identified 
as the Crum-Darboux transformation, is associated with the supercharges of the system.
In the current case, we will discuss briefly the technique in the context of the
first order, one-dimensional Dirac equation. We refer to  \cite{DiracDarboux} for 
more details. 

Let us have a physical system described by a solvable hermitian Hamiltonian 
\begin{equation}\label{seed}
h=i\sigma_2\partial_x+\Delta
\end{equation}
with real and symmetric matrix potential $\Delta=\Delta(x)$ and $x$ extending to the 
whole real axis. The physical eigenstates (solutions complying with prescribed boundary 
conditions) form a basis of the Hilbert space. Besides, the (formal) solutions of the 
stationary equation $hu=\lambda u$ are supposed to be known for any complex $\lambda$.  
 
We define the operator $L$ by 
\begin{equation}\label{L}
 L=U\frac{\partial}{\partial x} U^{-1}=\mathbf{1}\partial_x-U' U^{-1},
\end{equation}
where $U'=\partial U/\partial x$. The matrix $U=(u_1,u_2)$ is a chosen solution of the 
equation $h\, U=U\,\Lambda$ where the matrix $\Lambda=diag(\lambda_1,\lambda_2)$ has 
fixed real elements. The vectors  $u_{1(2)}$ satisfy $hu_{1(2)}=\lambda_{1(2)} u_{1(2)}$ 
and are chosen to be real. They form the kernel of $L$, $LU=0$ and do not need to be physical.
Next, we define the hermitian operator $\tilde{h}$ with the potential term explicitly 
dependent on $u_1$ and $u_2$ and corresponding eigenvalues $\lambda_1$ and $\lambda_2$,
\begin{eqnarray}\label{htilde}
 \tilde{h}&=&h+i[\sigma_2,U'U^{-1}]=\sigma_2h\sigma_2+\sigma_2[\sigma_2,U\Lambda U^{-1}]\nonumber\\
 &=&\sigma_2 h\sigma_2+\left(\frac{u_1^T\sigma_1u_2}{\det U}\sigma_3-\frac{u_1^T\sigma_3u_2}{\det U}\sigma_1\right)(\lambda_1-\lambda_2).\nonumber\\
\end{eqnarray}
We used here the identity
\begin{equation}\label{derid}
  \mathbf{1}\partial_{x}=-i\sigma_2(h-\Delta),
\end{equation}
which will be employed extensively in the following text. Notice that as long as $u_1$ 
and $u_2$ correspond to the same eigenvalue $\lambda_1=\lambda_2$, $\tilde{h}$ reduces 
to a unitary transformed seed Hamiltonian, $\tilde{h}=\sigma_2h\sigma_2$, for any $\Delta$. 
The Hamiltonians (\ref{seed}) and (\ref{htilde}) satisfy the following intertwining 
relations mediated by $L$ and $L^{\dagger}$,
\begin{equation}\label{intertwining}
 Lh=\tilde{h}L,\quad L^{\dagger}\tilde{h}=hL^{\dagger}.
\end{equation}
The conjugate operator $L^{\dagger}$ can be written as  $L^{\dagger}=-\partial_x+V'V^{-1}$ 
where $V=(U^{\dagger})^{-1}=(v_1,v_2)$. The columns $v_1$ and $v_2$ satisfy 
$L^{\dagger}v_{1(2)}=0$ and are solutions of $\tilde{h}v_{1(2)}=\lambda_{1(2)} v_{1(2)}$. 
There holds
\begin{equation}\nonumber
 \tilde{h}\, V=V\,\Lambda. 
\end{equation}

Each of the equations $(h-\lambda)\varphi=0$ and $(\tilde{h}-\lambda)\tilde{\varphi}=0$ 
has two independent formal solutions, let us denote them $\psi_{\lambda}$, $\xi_{\lambda}$ 
and $\tilde{\psi}_{\lambda}$, $\tilde{\xi}_{\lambda}$, respectively. 
For $\lambda\neq \lambda_{1(2)}$, the operators $L$ and $L^{\dagger}$ work as  one-to-one 
mappings between the two subspaces spanned by $\psi_{\lambda}$, $\xi_{\lambda}$ and 
$\tilde{\psi}_{\lambda}$, $\tilde{\xi}_{\lambda}$. They transform the (formal) eigenvectors 
of $h$ into the formal eigenvectors of $\tilde{h}$ and vice versa.

Let us consider now the four-dimensional subspace spanned by the solutions of 
$(h-\lambda_{1(2)})\varphi=0$. Two of the solutions, the vectors $u_1$ and $u_2$, compose 
the matrix $U$. We can use the other two vectors to define the matrix 
$\underline{U}=(\underline{u}_1,\underline{u}_2)$, which satisfies 
$h\underline{U}=\underline{U}\Lambda$ but is not annihilated by $L$. Similarly, we can define 
the matrix $\underline{V}=(\underline{v}_1,\underline{v}_2)$ from the solutions of 
$(\tilde{h}-\lambda_{1(2)})\tilde{\varphi}=0$ which satisfies 
$\tilde{h}\underline{V}=\underline{V}\Lambda$, but no linear combination of $\underline{v}_1$ 
and $\underline{v}_2$ is annihilated by $L^{\dagger}$. The intertwining operators then 
transform the matrices as $L\underline{U}\sim V$, $L^{\dagger}\underline{V}\sim U$. 
Hence, we get $L^{\dagger}LU=L^{\dagger}L\underline{U}=LL^{\dagger}V=LL^{\dagger}\underline{V}=0$. 
The latter equalities can be understood as the implication of the
alternative presentation for the products of the intertwining operators,
\begin{equation}\label{factorization}
 LL^{\dagger}=(\tilde{h}-\lambda_1)(\tilde{h}-\lambda_2),\quad L^{\dagger}L=(h-\lambda_1)(h-\lambda_2). 
\end{equation}
The spectrum of $\tilde{h}$ is identical with the spectrum of $h$ up to a possible difference 
in the energy levels $\lambda_1$ and/or $\lambda_2$. These energies are in the spectrum of 
either $h$ or $\tilde{h}$ if and only if the associated eigenvectors comply with the boundary 
conditions of the corresponding stationary equation. We will discuss specific examples where 
the spectrum of the new Hamiltonian $\tilde{h}$ contains additional discrete energies that are 
absent in the spectrum of $h$.
The eigenvector $\tilde{\phi}_{k}$ of $\tilde{h}$ corresponding to the energy level 
$\lambda_k\neq\lambda_1,\ \lambda_2$ can be expressed in terms of $L$ and  the eigenvectors 
$\phi_k$ of $h$ ($h\phi_{k}=\lambda_k\phi_k$), 
\begin{equation}\label{normalization}
 \tilde{\phi}_{k}=\frac{L\phi_{k}}{\sqrt{(\lambda_k-\lambda_1)(\lambda_k-\lambda_2)}}
,\quad  \tilde{h}\tilde{\phi}_{k}=\lambda_k\tilde{\phi}_{k}.\nonumber
\end{equation}
When defined in this way, the probability densities of $\tilde{\phi}_k$ and $\phi_k$ coincide. 

It is worth noticing that the system $\tilde{h}$ inherits integrals of motion of $h$. Indeed, if $S$ 
commutes with $h$, then the operator $\tilde{S}=LSL^{\dagger}$ generates a symmetry of $\tilde{h}$, 
$[\tilde{h},\tilde{S}]=0$. We will discuss this point in more detail in the context of the 
reflectionless models.

The potential term of $\tilde{h}$ in (\ref{htilde}) ceases to have a direct interpretation in the 
context of carbon nanotubes with the radial twist. As we are interested in the analysis of namely 
such systems, we require $\tilde{h}$ to be equivalent to the Hamiltonian in (\ref{eq1}); the 
term proportional to either $\sigma_1$ or $\sigma_3$  in (\ref{htilde}) should vanish. As these 
coefficients depend both on the potential of the seed Hamiltonian $h$ and  on its eigenvectors, 
it is rather difficult to meet this requirement in general. Instead, let us consider two special cases.

First, let us fix the initial Hamiltonian as 
\begin{equation}\label{hI}h_{I}=i\sigma_2\partial_x+m\sigma_3+\Delta_1\sigma_1,\end{equation} 
where $m>0$. We take $\lambda_1=m$ and $\lambda_2=0$ and denote $U_I\equiv U= (u_1,u_2)$ and 
$V_I\equiv V=(v_1,v_2)$, where explicitly $u_1=(u_{11},0)^T$, $u_2=(u_{12},u_{22})^T$ and
\begin{equation}      \label{UVI}                                                                                                                                                                                                                                                         
 U_I=\left(\begin{array}{cc}u_{11}&{u_{12}}\\0&u_{22}\end{array}\right),\quad  V_I=\left(\begin{array}{cc}\frac{1}{u_{11}}&0\\\frac{-u_{12}}{u_{11}u_{22}}& \frac{1}{u_{22}}\end{array}\right).                                                                                                                                                                                                                                                             \end{equation}
Comparison of the two matrices tells that if $u_1$ (or $u_2$) is a bound state of $h$, 
then $v_1$ (or $v_2$) cannot be bound state of $\tilde{h}$. Vice versa, if $v_1$ (or $v_2$) 
is a bound state of $\tilde{h}$, then $u_1$ (or $u_2$) is not normalizable. Using (\ref{htilde}) 
and $u^T_1\sigma_1u_2/\det U_I=1$, we get the Hamiltonian $\tilde{h}_I$, 
\begin{equation}\label{HI}
 \tilde{h}_I=i\sigma_2\partial_x-\left(\Delta_1+m\frac{u_{12}}{u_{22}}\right)\sigma_1,
\end{equation}
with the required form of the potential.  

In the second case, we take the seed Hamiltonian as 
\begin{equation}\label{hII}h_{II}=i\sigma_2\partial_x+(\Delta_1+m)\sigma_1\end{equation}
and fix $\lambda_1=-\lambda_2>0$. The vectors $u_{1(2)}$ are chosen as $u_1=(u_{11},u_{21})^T$ 
and $u_2=\sigma_3u_1$. They satisfy $u_1^T\sigma_1u_2=0$. The matrices $U_{II}\equiv U$ and 
$V_{II}=V$ are in this case
\begin{equation}\label{UVII}                                                                                                                                                                                                                                                               
 U_{II}=\left(\begin{array}{cc}u_{11}&{u_{11}}\\u_{21}&-u_{21}\end{array}\right),\quad  V_{II}=\frac{1}{2}\left(\begin{array}{cc}{u_{11}^{-1}}&{u_{11}^{-1}}\\{u_{21}^{-1}}& -{u_{21}^{-1}}\end{array}\right)                                                                                                                                                                                                                                                       \end{equation}
and the Hamiltonian (\ref{htilde}) acquires the form
\begin{equation}\label{HII}
 \tilde{h}_{II}=i\sigma_2\partial_x-\left(\Delta_1+m-\lambda_1\frac{u_{11}^2+u_{21}^2}{u_{11}u_{21}}\right)\sigma_1.
\end{equation}  
We can deduce that if $u_1$ is a bound state of $h$, so is the vector $u_2$ and neither $v_1$ 
or $v_2$ can be normalized. Vice versa, if $v_1$ and $v_2$ are bound states of $\tilde{h}$, 
the vectors $u_1$ and $u_2$ are not normalizable.

In the end of the section, let us notice that there is an alternative interpretation in 
dealing with the intertwining relations and the involved operators. Inspired by the
SUSY quantum mechanics, we can define the extended, first order matrix operators
\begin{equation}
 \mathcal{H}=\left(\begin{array}{cc}\tilde{h}&0\\0&h\end{array}\right),\quad \mathcal{Q}_1=\left(\begin{array}{cc}0&L\\L^{\dagger}&0\end{array}\right),\quad \mathcal{Q}_2=i\left(\begin{array}{cc}0&L\\-L^{\dagger}&0\end{array}\right),
\end{equation}
which establish the N = 2 (nonlinear) supersymmetry.  The grading operator 
$\Gamma=\mbox{diag}({\bf 1},-{\bf 1})$ classifies the Hamiltonian $\mathcal{H}$ 
as bosonic ($[\mathcal{H},\Gamma]=0$), while both $\mathcal{Q}_1$ and 
$\mathcal{Q}_2=i\Gamma\mathcal{Q}_1$ are fermionic, $\{\mathcal{Q}_a,\Gamma\}=0$ for $a=1,2.$ 

Contrary to the SUSY quantum mechanics based on the second order matrix Hamiltonian,
here  both the Hamiltonian $\mathcal{H}$ and the supercharges
$\mathcal{Q}_{1,2}$ are the first order differential operators.  The associated 
(nonlinear) superalgebra
\begin{equation}\label{susyextend}
[\mathcal{H},\mathcal{Q}_a]=0,\quad \{\mathcal{Q}_a,\mathcal{Q}_b\}=2\delta_{ab}(\mathcal{H}-\lambda_1)(\mathcal{H}-\lambda_2)
\end{equation}
encodes the intertwining relations (\ref{intertwining}) together with the factorization 
(\ref{factorization}).

\section{Green's function and LDOS for the twisted nanotubes}

We shall derive formula for the Green's function of $\tilde{h}$ in terms of the intertwining 
operator $L$ and the Green's function of the initial Hamiltonian $h$. In the end of the section, 
we will discuss the explicit form of the LDOS for the systems described by $\tilde{h}_I$ and 
$\tilde{h}_{II}$ of the form (\ref{HI}) and (\ref{HII}) corresponding to the twisted carbon nanotubes.

Let us start with the hermitian Hamiltonian $h=i\sigma_2\partial_x+\Delta(x)$. The potential term is 
required to be real and symmetric. The (generalized) eigenstates $\phi_{\lambda}$ of $h$ have to 
satisfy the following boundary conditions
\begin{equation}\label{heigen}
h\phi_{\lambda}=\lambda\phi_{\lambda},\quad \phi_{\lambda}(x)|_{x\rightarrow\pm\infty}\sim f_{\pm}(\lambda,x),\quad \lambda\in\mathbb{R}.
\end{equation} 
The symbol $\sim$ means here that the elements of the eigenvector $\phi_{\lambda}$ are proportional 
asymptotically to the function $f_{\pm}(x,\lambda)$. We prefer to leave the boundary conditions 
unspecified explicitly at the moment. 
They will be discussed for the reflectionless models later in the text.

The Green's function associated with the Hamiltonian $h$ is defined as a solution of the equation
\begin{equation}\label{Gdefiningeq}
 (h-\lambda)G(x,y;\lambda)=\delta(x-y),\quad \lambda\in \mathbb{C}.
\end{equation}
It has to satisfy the same boundary conditions as the eigenstates of $h$, i.e. the matrix 
elements of the Green's function are  proportional to $f_{\pm}(\lambda,x)$ in the limit 
${x\rightarrow \pm\infty}$. Being effectively the inverse of $(h-\lambda)$, the Green's 
function is not well defined for $\lambda$ from the spectrum $\sigma(h)$ of $h$. It has  
simple poles for $\lambda$ corresponding to discrete energies.
If $\lambda$ is in the continuous spectrum, then we can find the limit 
$G^{\pm}(x,y;\lambda)=\lim_{\eta\rightarrow0}G(x,y;\lambda\pm i\eta)$, see e.g. \cite{economou}.  

The differential equation in (\ref{heigen}) has two formal independent solutions 
 $\psi_{\lambda}(x)$ and $\xi_{\lambda}(x)$ for any $\lambda\in\mathbb{C}$.
For $\lambda\notin \sigma(h)$, we can fix $\psi_{\lambda}$ and $\xi_{\lambda}$ such that 
each of the functions complies with the boundary condition in one of the boundaries; i.e. 
we fix $\psi_{\lambda}(x)|_{x\rightarrow +\infty}\sim f_{+}(x,\lambda)$ and 
$\xi_{\lambda}(x)|_{x\rightarrow -\infty}\sim f_{-}(x,\lambda)$. 
These functions can be employed in the construction of the Green's function in the following way
\begin{equation}\label{GreenG}G(x,y;\lambda)=\frac{\psi_{\lambda}(x)\xi_{\lambda}(y)^T\theta(x-y)+\xi_{\lambda}(x)\psi_{\lambda}(y)^T\theta(y-x)}{W(\psi_{\lambda},\xi_{\lambda})},\end{equation}
where $\theta$ is the step function. The quantity
\begin{equation}
W(\psi,\xi)=i\psi(x)^T\sigma_2\xi(x)
\end{equation}
is the analog of Wronskian for Dirac equation. It is constant for two independent 
solutions $\psi_{\lambda}$ and $\xi_{\lambda}$ corresponding to the eigenvalue 
$\lambda$ of $h$. Indeed, direct calculation with the use of (\ref{derid}) shows that 
$\partial_xW(\psi,\xi)=0$. The Green's function defined in (\ref{GreenG}) then solves 
(\ref{Gdefiningeq}) and manifestly satisfies the prescribed boundary conditions for $x\rightarrow\pm\infty$. 

Let us pass to the system described by $\tilde{h}$ and construct its Green's function with 
the use of (\ref{GreenG}). We suppose that $L$ transforms appropriately the boundary 
conditions associated with $h$ to the boundary conditions prescribed for the eigenstates 
of $\tilde{h}$. We can define the functions 
$\tilde{\psi}_{\lambda}=\frac{L\psi_{\lambda}}{\sqrt{(\lambda-\lambda_1)(\lambda-\lambda_2)}}$ and 
$\tilde{\xi}_{\lambda}=\frac{L\xi_{\lambda}}{\sqrt{(\lambda-\lambda_1)(\lambda-\lambda_2)}}$. 
They solve $\tilde{h}\tilde{\psi}_{\lambda}=\lambda\tilde{\psi}_{\lambda}$, 
$\tilde{h}\tilde{\xi}_{\lambda}=\lambda\tilde{\xi}_{\lambda}$  
and satisfy the prescribed boundary condition in $+\infty$ or $-\infty$, respectively. 
The Green's function associated with $\tilde{h}$ can be written then as
\begin{eqnarray}\label{GreentildeG}\tilde{G}(x,y;\lambda)&=&\frac{\tilde{\psi}_{\lambda}(x)\tilde{\xi}_{\lambda}(y)^T\theta(x-y)+\tilde{\xi}_{\lambda}(x)\tilde{\psi}_{\lambda}(y)^T\theta(y-x)}{W(\tilde{\psi}_{\lambda},\tilde{\xi}_{\lambda})}\nonumber\\
 &=&\frac{1}{W(\psi_{\lambda},\xi_{\lambda})}\left[\frac{(L\psi_{\lambda})(x)(L\xi_{\lambda})^T(y)\theta(x-y)}{(\lambda-\lambda_1)(\lambda-\lambda_2)}\right.\nonumber\\
&&\left.+\frac{(L\xi_{\lambda})(x)(L\psi_{\lambda})^T(y)\theta(y-x)}{(\lambda-\lambda_1)(\lambda-\lambda_2)}\right].
\end{eqnarray}
We used the fact that the Wronskian is invariant with respect to the Darboux transformation 
(\ref{L}), $W(\tilde{\psi}_{\lambda},\tilde{\xi}_{\lambda})=W(\psi_{\lambda},\xi_{\lambda})$. 
We refer to \cite{Dunne} or \cite{halberg} where the proof of this relation can be found. 
Let us mention that the a different supersymmetric approach to Green's 
functions of Dirac operators was examined in \cite{Feinbergsusy} where a modification of the 
standard supersymmetry (based on second-order Hamiltonians) was discussed.

The eigenvectors of $\tilde{h}$ can be written as
\begin{equation}\label{algebraicL}
 \tilde{\psi}_{\lambda}={\cal L}(\lambda,x)\psi_{\lambda},\quad {\cal L}(\lambda,x)=-i\sigma_2\frac{\lambda-U(x)\Lambda U^{-1}(x)}{\sqrt{(\lambda-\lambda_1)(\lambda-\lambda_2)}},
\end{equation}
where we used (\ref{derid}) again.
This allows us to rewrite the Green's
function (\ref{GreentildeG}) in particularly simple form
\begin{equation}\label{LGL}
 \tilde{G}(x,y;\lambda)={\mathcal L}(\lambda,x)G(x,y;\lambda){\mathcal L}^T(\lambda,y).
\end{equation}
Hence, $\tilde{G}(x,y;\lambda)$ can be obtained by purely algebraic means without the use of 
any differential operator; it can be obtained just by multiplication of $G(x,y;\lambda)$ with 
simple matrix operators (\ref{algebraicL}).  

The local density of states $\rho(x,\lambda)$ associated with $h$ is computed in the following manner  
\begin{eqnarray}\label{LDOS}
 \rho(x,\lambda)&=&-\frac{1}{\pi}\lim_{\mbox{Im}\lambda\rightarrow 0_+}\mbox{Im}\,Tr \,G(x,x;\lambda),
\end{eqnarray}
where the trace is taken over the matrix degrees of freedom. Using (\ref{LGL}), we can write 
LDOS $\tilde{\rho}$ for $\tilde{h}$ as
\begin{equation}\label{tildeLDOS}
 \tilde{\rho}(x,\lambda)=-\frac{1}{\pi}\lim_{\mbox{Im}\lambda\rightarrow 0_+}\mbox{Im}\, Tr\left(\mathcal{L}(\lambda,x)^T\mathcal{L}(\lambda,x)\,G(x,x;\lambda)\right).
\end{equation}
Notice that the formulas (\ref{LGL}) and (\ref{tildeLDOS}) are valid for a general class of 
the seed Hamiltonians with real and symmetric potential. 

In the literature (see, e.g. \cite{Dunne}, \cite{Feinberg}, \cite{Kos}), the operator 
$G(x,x;\lambda)$ is called Gorkov Green's function 
or diagonal resolvent of $h$. The
Green's function of the Schr\"odinger operators and generalized Sturm-Liouville equation was 
studied in \cite{greenSamsonov} and \cite{halberg2} in the context of intertwining relations.

We turn our attention to the systems represented by $\tilde{h}_I$ and $\tilde{h}_{II}$ which 
describe the carbon nanotubes with the radial twist. It is supposed that the Green's functions 
of both $h_I$ and $h_{II}$ are known. We denote them $G_I(x,y;\lambda)$ and $G_{II}(x,y;\lambda)$.  
The operators $\mathcal{L}_I(\lambda,x)$ and $\mathcal{L}_{II}(\lambda,x)$ based on $U_{I}$ and 
$U_{II}$ respectively acquire particularly simple form
\begin{equation}\nonumber
\mathcal{L}_I(\lambda,x)= \frac{1}{\sqrt{\lambda(\lambda-m)}}\left(\begin{array}{cc}0&-\lambda\\-m+\lambda&m\frac{u_{12}}{u_{22}}\end{array}\right)
\end{equation}
and 
\begin{equation}\nonumber
 \mathcal{L}_{II}(\lambda,x)= \frac{1}{\sqrt{(\lambda^2-\lambda_1^2)}}\left(\begin{array}{cc}\lambda_1\frac{u_{21}}{u_{11}}&-\lambda\\\lambda&-\lambda_1\frac{u_{11}}{u_{21}}\end{array}\right).
\end{equation}
The trace of the $\tilde{G}_{I}$ can be computed directly in terms of the vectors $u_1$ and $u_2$. 
A straightforward computation gives
\begin{eqnarray}
\mbox{Tr}(\tilde{G}_{I}(x,x;\lambda))&=&g_0-\frac{m}{\lambda-m}g_3+\frac{m^2\,g_0(u_1^{\dagger}u_1)(u_2^{\dagger}u_2)}{2\lambda(\lambda-m)(\det U_I)^2}\nonumber
\end{eqnarray}
\begin{equation}\label{GGI}
 +\frac{m^2u_1^{\dagger}u_1}{2\lambda(\lambda-m)(\det U_I)^2}\left(-g_3u_2^{\dagger}\sigma_3u_2+g_1\frac{\lambda-m}{m}u_2^{\dagger}\sigma_1u_2\right).
\end{equation}
Here we used the abbreviated notation $g_0=\mbox{Tr}\,G_I(x,x;\lambda)$ and 
$g_j=\mbox{Tr}(\sigma_jG_I(x,x;\lambda))$ for $j=1,3$.
We can obtain similar expression for the trace of the $\tilde{G}_{II}(x,x;\lambda)$:
\begin{eqnarray}
\mbox{Tr}(\tilde{G}_{II}(x,x;\lambda))&=&g_0+\frac{2\lambda_1^2g_0(u_1^{\dagger}u_1)^2}{(\lambda^2-\lambda_1^2)(\det U_{II})^2}\nonumber
\end{eqnarray}
\begin{equation}\label{GGII}
 +\frac{2\lambda_1^2u_1^{\dagger}u_1}{(\lambda^2-\lambda_1^2)(\det U_{II})^2}\left(-g_3u_1^{\dagger}\sigma_3u_1-g_1\frac{\lambda}{\lambda_1}u_1^{\dagger}\sigma_1u_1\right).
\end{equation}
The notation used here is like in (\ref{GGI}) with the replacement of  
${G}_{I}(x,x;\lambda)$ by ${G}_{II}(x,x;\lambda)$.

\section{\label{reflectionless}Perfect tunneling in the twisted carbon nanotubes}

There exists an exceptional class of exactly solvable systems whose Hamiltonian 
$\tilde{h}$ is intertwined with the Hamiltonian of the free particle. The peculiar 
and simple properties of the latter model are manifested in these systems as well. 
In particular, they share the trivial scattering characteristics of the interaction-free 
model, i.e. they are \textit{reflectionless}. The eigenstates of both the free-particle 
system and the reflectionless models are 
subject to the same boundary conditions; the scattering states have to be oscillating 
in the infinity while the bound states should decay exponentially for $|x|\rightarrow\infty$. 
Additionally, the reflectionless systems inherit the integral
of motion that in the free particle system plays
the role of generator of translations.

The stationary equation $h\phi=\lambda\phi$, where 
$h=i\sigma_2\partial_x+m\sigma_3$,
is translationally invariant, i.e. the Hamiltonian commutes with $p=-i\partial_x$. 
We can find the common eigenstates of $h$ and $p$. 
The latter operator distinguishes the two scattering states corresponding to each 
doubly degenerate energy level. It annihilates the singlet states $u_{+}=(1,0)^T$ 
and $u_-=(0,1)^T$ that correspond to the edges $\lambda=\pm m$ of the positive and 
negative part of the continuous spectrum (which are called the conduction and the 
valence band respectively in the context of nanotubes). The involved operators close 
the nonlinear superalgebra 
\begin{equation}\label{freehiddensusy}
[p,h]=0,\quad  \{p,p\}=2(h-m)(h+m),  
\end{equation}
which is graded by the parity operator $\Gamma=R\sigma_3$ ($RxR=-x$, $\Gamma^2=1$).
Let us stress that this supersymmetric structure is completely different from (\ref{susyextend}). 
In this case, the supersymmetry is rather hidden; the two fold degeneracy of energy levels, 
distinguished by the integral of motion $p$, emerges within the spectrum of the unextended
Hamiltonian $h$.  

The Hamiltonian $\tilde{h}$ inherits a modified version of the nontrivial integral of 
motion $p$. It can be found by dressing of the initial symmetry operator,
\begin{equation}\label{A}
 \tilde{p}=L\, p\,L^{\dagger},\quad [\tilde{p},\tilde{h}]=0.
\end{equation}
It annihilates the states $v_1$ and $v_2$ together with the vectors $\tilde{v}_{\pm}$ 
which are defined as $\tilde{v}_{\pm}=Lu_{\pm}$. The operator $\tilde{p}$, like 
$p$ in the free particle model, reflects the degeneracy 
of the spectrum; it can distinguish the scattering states corresponding to the same energy 
level.  
The superalgebra (\ref{freehiddensusy}) can be recovered in the modified form
\begin{equation}\label{dressedhiddensusy}
 [\tilde{p},\tilde{h}]=0,\quad \{\tilde{p},\tilde{p}\}=2(\tilde{h}^2-m^2)(\tilde{h}-\lambda_1)^2(\tilde{h}-\lambda_2)^2.
\end{equation}
Hence, the square of $\tilde{p}$  is the spectral polynomial of $\tilde{h}$. 
It is worth noticing that the same algebraic structure, the hidden supersymmetry, 
was discussed in detail for both relativistic and nonrelativistic finite-gap systems 
in \cite{BdG}, \cite{hiddensusy}, \cite{mirror}, \cite{AdS2}. In this context, the 
integral $\tilde{p}$ can be identified as the Lax operator of the system represented by $\tilde{h}$.

\subsection{Single-kink system }
The first model will be derived with the use of the seed Hamiltonian $h_I$ in (\ref{hI})
with $\Delta_1(x)=0$,
\begin{equation}\nonumber
 h_I=i\sigma_2\partial_x+m\sigma_3.
\end{equation}
We will compute its LDOS and discuss the realization of the parity operator of the hidden supersymmetry.

We require that the new Hamiltonian $\tilde{h}$ has a single bound state with zero energy. 
To meet this requirement, we fix the matrix $U_I$ as
\begin{equation}\nonumber
 U_I=\sqrt\frac{2}{m}\left(\begin{array}{cc}1&-\sinh mx\\0&\cosh mx\end{array}\right),  
\end{equation}
and the intertwining operator $L$ as
\begin{equation}\label{LI}
  L_I=\mathbf{1}\,\partial_x+m\left(\begin{array}{cc}0&1\\0&-\tanh mx\end{array}\right). 
\end{equation}
The explicit form of the matrix $V_I$ is then
\begin{equation}V_I=\sqrt{\frac{m}{2}}\left(\begin{array}{cc}1&0\\\tanh\, mx&\mbox{sech}\, mx\end{array}\right). \nonumber                                                                                                                                                                                                                                                                                                                                                                                                                                        \end{equation} 
The associated Hamiltonian $\tilde{h}_I$ then reads
\begin{equation}\label{hIex}
 \tilde{h}_I=i\sigma_2\partial_x+m\,\sigma_1\,\tanh mx.
\end{equation}  
The operator $\tilde{h}_I$ has the normalized bound state $v_2$
\begin{equation}\nonumber
  v_2=\left(0,\sqrt{\frac{m}{2}}\,\mbox{sech}\, mx\right)^T.                                                                                                                                                                                                                                                
\end{equation}
Let us notice that the operator (\ref{hIex}) appears in description of many physical systems, 
e.g. in the continuum model for solitons in polyacetylene \cite{Takayama} or in the analysis 
of the static fermionic bags of the Gross-Neveu model \cite{Feinberg}. 

We can use (\ref{LDOS}) together with (\ref{GGI}) to compute the LDOS of the system.  
It acquires the following simple form
\begin{equation}\label{exILDOS}
 \tilde{\rho}_I(x,\lambda)=\frac{2|\lambda|^2-m^2\mbox{sech}^2 mx}{2\pi|\lambda|\sqrt{|\lambda^2-m^2|}}\theta(\lambda^2-m^2).
\end{equation}
The presence of the step function $\theta$ reflects that fact that imaginary part of 
(\ref{GGI}) for $|\lambda|<m$ is zero and, hence, $\rho(x,\lambda)$ vanishes identically.
The formula (\ref{exILDOS}) can be rewritten with the use of the LDOS of the free 
particle \begin{equation}\nonumber\rho_I(x,\lambda)=\frac{|\lambda|}{\pi\sqrt{|\lambda^2-m^2|}}\,\theta(\lambda^2-m^2)\end{equation}
and the density of probability of the bound state $v_2$,
\begin{equation}\label{LDOSI}
 \tilde{\rho}_{I}(x,\lambda)=\rho_I(x,\lambda)\left(1-\frac{m}{|\lambda|^2}v_2^{\dagger}v_2\right).
\end{equation}
The coefficient of the second term is just the difference of the densities of states of $h$ and $\tilde{h}$,
\begin{equation}\nonumber
 \int_{\mathbb{R}}(\rho_I-\tilde{\rho}_I)dx=\frac{m\,\theta(\lambda^2-m^2)}{\pi|\lambda|\sqrt{|\lambda^2-m^2|}}.
\end{equation}
Let us notice that the difference of densities of states for Dirac particle on the \textit{finite} interval 
with Dirichlet boundary conditions was discussed in \cite{halberg}.

The hidden superalgebra (\ref{dressedhiddensusy}), closed by $\tilde{h}_I$ and $\tilde{p}_I=L_IpL^{\dagger}_I$, 
reads explicitly
\begin{equation}\nonumber
 [\tilde{h}_I,\tilde{p}_I]=0,\quad \{\tilde{p}_I,\tilde{p}_I\}=2(\tilde{h}_I-m)^3\tilde{h}_I^2(\tilde{h}_I+m). 
\end{equation}
The parity operator $\tilde{\Gamma}=R\, \sigma_3$, $\tilde{\Gamma}^2=1$, 
classifies $\tilde{h}_I$ and $\tilde{p}_I$ as, respectively, bosonic and fermionic operators, 
$[\tilde{h}_I,\tilde{\Gamma}]=\{\tilde{p}_I,\tilde{\Gamma}\}=0$. 

The potential in (\ref{hIex}) can be associated with the displacement vector
$\mathbf{d}=(0,\ln\cosh mx).$
The corresponding twist of the metallic nanotube is illustrated in Figure \ref{bubu}.  
Hence, the nanotube is twisted in one direction up to the center (origin)
where the orientation of the twist gets changed.

\newsavebox{\figPT}        % vytvoření boxu
    \savebox{\figPT}{          % uložení obrázku do boxu
    \scalebox{1}{
    \includegraphics[scale=.7]{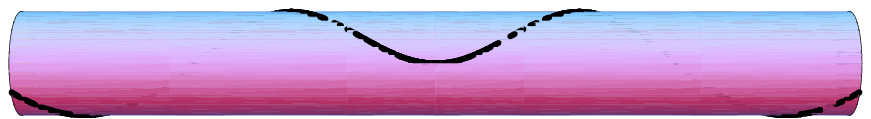}
    }
    }
\begin{figure}\begin{center}
\usebox{\figPT}\caption{\label{bubu}The metallic nanotube with the twist associated with $ d_y\sim\ln\cosh mx$ 
and the Hamiltonian (\ref{hIex}). In the untwisted nanotube, the black line would be straight.}
\end{center}\end{figure}

\subsection{Double-kink model}
Here we construct the system with two bound states. We shall employ the scheme 
discussed in (\ref{hII})-(\ref{HII}). Fixing $\Delta_1(x)=0$ in (\ref{hII}), we get 
the free particle Hamiltonian 
\begin{equation}\nonumber
 h_{II}=i\sigma_2\partial_x+m\sigma_1.
\end{equation}
We choose the components of $U_{II}$ as
\begin{equation}\nonumber
 u_{11}=\frac{1}{\sqrt{k}}\cosh k x,\quad u_{21}=\frac{1}{\sqrt{k}}\cosh (k x+a)
\end{equation}
where
\begin{equation}\nonumber
 a=\frac{1}{2}\log\frac{m-k}{m+k}, \quad k=\sqrt{m^2-\lambda_1^2},\quad 0<\lambda_1<m.
\end{equation}
The intertwining operator acquires a diagonal form
\begin{equation}\label{Lreflectionless}
 L_{II}=\mathbf{1}\,\partial_x-k\left(\begin{array}{cc}\tanh(kx)&0\\0&\tanh (kx+a)\end{array}\right).\end{equation}
The formula (\ref{HII}) then provides the explicit form of the Hamiltonian $\tilde{h}_{II}$                                                               
\begin{eqnarray}\label{hIIex}
\tilde{h}_{II}&=&i\sigma_2\partial_x+\left(-m+\lambda_1\frac{\cosh^2 kx+\cosh^2(kx+a)}{\cosh kx\cosh(kx+a)}\right)\sigma_1\nonumber\\
 &=&i\sigma_2\partial_x+(m-k\tanh k x+k\tanh(k x +a))\sigma_1.
\end{eqnarray}
The potential term is asymptotically equal to $m\sigma_1$. The system has two bound states 
represented by the normalized vectors $v_1$ and $v_2=\sigma_3\,v_1$ where 
\begin{equation}\nonumber
v_1=\frac{\sqrt{k}}{2}(\mbox{sech} kx,\mbox{sech}(kx+a))^T.
\end{equation} 
Notice that the equation (\ref{hIIex}) appeared in the analysis of the Dashen, Hasslacher, and Neveu
kink-antikink baryons in Gross-Neveu model \cite{kink-antikink}.

The local density of states in the current system can be computed directly with the 
use of (\ref{GGII}). We get
\begin{eqnarray}\nonumber
 \tilde{\rho}_{II}(x,\lambda)&=&\frac{|\lambda|\left(1-\frac{k^2}{2(\lambda^2-\lambda_1^2)}(\mbox{sech}^2kx+\mbox{sech}^2(kx+a))\right)}{\pi\sqrt{|m^2-\lambda^2|}}\nonumber\\&&\times\theta(\lambda^2-m^2).
\end{eqnarray}
Likewise in the preceding example, it can be written as the LDOS of the free particle corrected 
by the term proportional to the probability density of the bound states, 
\begin{equation}\label{LDOSII}
 \tilde{\rho}_{II}(x,\lambda)=\rho_{II}(x,\lambda)\left(1-\frac{2\,k\,v_1^{\dagger}v_1}{(\lambda^2-\lambda_1^2)}\right),
\end{equation}
where $\rho_{II}(x,\lambda)=\rho_{I}(x,\lambda)$.
This time, the difference of the densities of states is
\begin{equation}\nonumber
 \Delta DOS=\int_{\mathbb{R}}(\rho_0-\rho_1)dx=\frac{2k\,|\lambda|\theta(\lambda^2-m^2)}{\pi\sqrt{|m^2-\lambda^2|}(\lambda^2-\lambda_1^2)}.
\end{equation}

The hidden superalgebra associated with the system, 
\begin{equation}
 [\tilde{h}_{II},\tilde{p}_{II}]=0,\quad
\{\tilde{p}_{II},\tilde{p}_{II}\}=2(\tilde{h}_{II}^2-m^2)(\tilde{h}_{II}^2-\lambda_1^2)^2,\nonumber\end{equation}
is graded by the operator
$\tilde{\Gamma}=R\, R_{\alpha} \sigma_1$ where $ R_{\alpha}f(x)=f(x+\alpha)R_{\alpha}
$, $R_{\alpha}R=RR_{-\alpha}=R(R_{\alpha})^{-1}$ and $\alpha=-\frac{a}{k}$.
This grading operator (represented in another form) was also discussed 
in \cite{mirror}.

The vector potential in (\ref{hIIex}) corresponds to the displacement
$
 d_y=mx-\ln\cosh kx+\ln\cosh(kx+a).
$
The corresponding twist of the metallic nanotube does not change its orientation asymptotically, 
see Figure \ref{fig3} for illustration.
\newsavebox{\figdva}        % vytvoření boxu
    \savebox{\figdva}{          % uložení obrázku do boxu
    \scalebox{1}{
    \includegraphics[scale=.65]{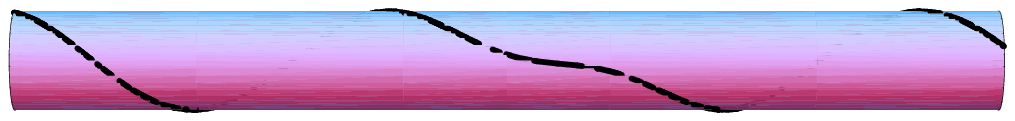}
    }
    }
 
\begin{figure}\begin{center}
\usebox{\figdva}\caption{\label{fig3}The metallic nanotube with the twist associated with 
$ d_y\sim mx+\ln\frac{\cosh(kx+a)}{\cosh kx}$ 
and the Hamiltonian (\ref{hIIex}).}
\end{center}\end{figure}
 
We can find another physically interesting setting described by $\tilde{h}_{II}$. 
We can divide the vector potential in (\ref{hIIex}) into two parts. The 
first part is associated with the asymptotically vanishing twist of the nanotube, 
$\tilde{\Delta}_T=-k\tanh k x+k\tanh(k x +a)$.  The second part is constant, 
$\tilde{\Delta}_{MG}=m$, and corresponds to the homogeneous external 
magnetic field which is parallel with the axis of the nanotube. Hence, $\tilde{h}_{II}$ 
describes the metallic nanotube which is asymptotically free of twists, however, the external 
constant magnetic field is present. See Figure \ref{fig4} for illustration. 
\newsavebox{\figctyri}        % vytvoření boxu
    \savebox{\figctyri}{          % uložení obrázku do boxu
    \scalebox{1}{
    \includegraphics[scale=.65]{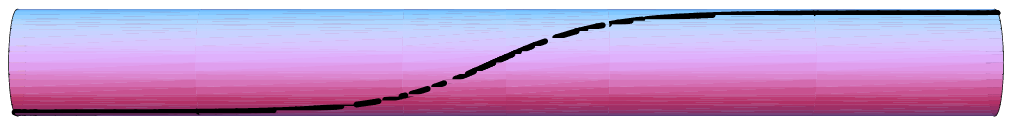}
    }
    }
 
\begin{figure}\begin{center}
\usebox{\figctyri}\caption{\label{fig4}The nanotube associated with the Hamiltonian 
(\ref{hIIex}) and the twist corresponding to (\ref{twistII}).
The constant part of the magnetic field in (\ref{hIIex}), $\tilde{\Delta}_{MG}=m$, can be attributed to 
the external magnetic field or to the semi-conducting character of the nanotube.}
\end{center}\end{figure}
The uniform external field opens a gap of the width $2m$ in the spectrum while the 
asymptotically vanishing twist induces two bound states in the gap. The model allows 
to compute the bound state energies as a function of an asymptotic (global) twist. 
Indeed, the twist associated with $\tilde{\Delta}_T $ is 
\begin{equation}\label{twistII}
  d_y= \ln\frac{\cosh(kx+a)}{\cosh kx}.
\end{equation}
The asymptotic twist corresponds to 
\begin{equation}\label{asymptotictwist}
 \delta d=|\lim_{x\rightarrow\infty}d_y-\lim_{x\rightarrow-\infty}d_y\,|=2|a|=-\ln\frac{m-\sqrt{m^2-\lambda_1^2}}{m+\sqrt{m^2-\lambda_1^2}}. 
\end{equation}
The dependence of the bound state energies on $\delta d$ then acquires the following simple form
\begin{equation}\label{induced}
 \lambda_1=\pm 2m \frac{e^{\frac{\delta d}{2}}}{1+e^{\delta d}}
\end{equation}
and is plotted in Figure \ref{fig5}.

\newsavebox{\figpet}        % vytvoření boxu
    \savebox{\figpet}{          % uložení obrázku do boxu
    \scalebox{1}{
    \includegraphics[scale=1.15]{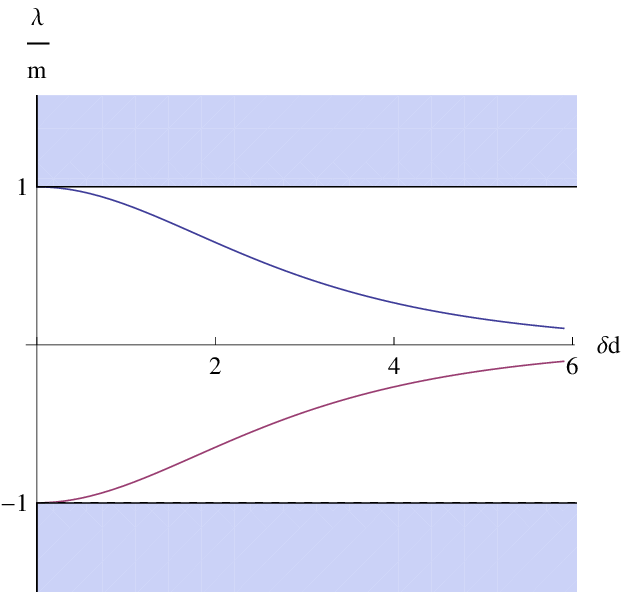}
    }
    }
 
\begin{figure}\begin{center}
\usebox{\figpet}\caption{\label{fig5}The spectrum of the Hamiltonian (\ref{hIIex}). 
The asymptotic twist of the nanotube  (\ref{asymptotictwist}) induces bound states of 
energies (\ref{induced}). The parameter $m$ is proportional to the inverse of the radius 
of the nanotube, see \cite{scale}.}
\end{center}\end{figure}

Up to now, the twisted nanotubes were considered to be metallic. The analysis can be 
extended to semi-conducting nanotubes without any difficulties; a constant, nonzero, 
part of the vector potential $\tilde{\Delta}_1$ has to be associated with the internal 
characteristics (the orientation of the hexagonal lattice) of the nanotube. 
Let us notice in this context that the metallic nanotube can be converted into the 
semi-conducting one just by switching on the constant magnetic flux parallel to the 
axis of the nanotube. This fact was experimentally confirmed in \cite{ABoscillations} 
and coined as Aharonov-Bohm oscillations of the carbon nanotubes. Turning back to (\ref{hIIex}), 
we can interpret the Hamiltonian as the energy operator of the semiconducting nanotube 
with a radial twist associated to $\tilde{\Delta}_{T}$. The constant part 
$p_y\equiv\tilde{\Delta}_{MG}$ of the potential appears due to the semiconducting nature of the nanotube.

\section{Discussion}

The expressions (\ref{LDOSI}) and (\ref{LDOSII}) can be written in the unified form
\begin{equation}\label{obecny}\nonumber
 \tilde{\rho}_{I(II)}=\rho_{I(II)}\left(1-\sum_j\frac{\sqrt{m^2-\lambda_j^2}}{(\lambda^2-\lambda_j^2)}v_j^{\dagger}v_j\right),
\end{equation}
where the sum is taken over the \textit{normalized} bound states of $\tilde{h}_{I(II)}$ 
annihilated by $L_{I(II)}$. It manifests a decrease of the LDOS in the regions where the 
bound states are localized. However, it is rather just a peculiar property of the 
discussed reflectionless models \cite{feinbergformule}. In general case, the LDOS (\ref{GGII}) of $\tilde{h}_{II}$ 
acquires  the following form in terms of the vectors $v_1$ and $v_2$,
\begin{eqnarray}
\mbox{Tr}(\tilde{G}_{II}(x,x;\lambda))&=&\mbox{Tr}({G}_{II}(x,x;\lambda))\nonumber
\end{eqnarray}
\begin{equation}
 +\frac{2\lambda_1^2v_1^{\dagger}v_1\left(g_0v_1^{\dagger}v_1+g_3v_1^{\dagger}\sigma_3v_1-g_1\frac{\lambda}{\lambda_1}v_1^{\dagger}\sigma_1v_1\right)}{(\lambda^2-\lambda_1^2)(\det V)^2}.\nonumber
\end{equation}
When $h_{II}$ is equal to the free particle Hamiltonian, the coefficient of $v_1^{\dagger}v_1$ 
reduces to a constant. Nevertheless, this apparently does not hold in the general case.

We restricted our consideration just to the systems described by the 
Hamiltonian $\tilde{h}=i\sigma_2\partial_x+\Delta_1(x)\sigma_1$. However,  the potential 
term of the seed Hamiltonian $h$ in (\ref{seed}) can acquire quite generic form, yet 
keeping valid the formulas (\ref{LGL}) and (\ref{tildeLDOS}) for the Green's functions and  
for the LDOS. Other results are more sensitive to the explicit form of the potential. 
The term $\Delta_2(x)\sigma_2$ cannot cause any substantial modifications; it 
would play just the role of non-physical gauge field. In contrary, impact of the diagonal term 
$\mathbf{1}\Delta_0+\Delta_3(x)\sigma_3$ in $\tilde{h}$ would be much deeper: in general, it would break the 
symmetry $\sigma(\tilde{h})=-\sigma(\tilde{h})$ of the spectrum $\sigma(\tilde{h})$ 
of $\tilde{h}$. In the context of Dirac particles in the carbon nanotubes, the potential $\Delta_3(x)\sigma_3$ would 
have different sign for the spin -up and -down components of wave function, i.e. this potential would change 
the sign on the two sublattices that form the crystal. Physical realization of such a scenario 
in the considered condensed matter system is not clear.  It is remarkable that the 
Darboux transformation (\ref{L}) does not alter the 
form of $\Delta_0$; the new Hamiltonian $\tilde{h}$ shares the same electrostatic potential 
as the seed Hamiltonian. We notice that the electrostatic potential can be also 
altered via the so-called 0-th order supersymmetry, as it was discussed in \cite{nasKlein}. 

In the discussed systems represented by the stationary equation (\ref{eq1}), 
the analysis of the bound states can be facilitated by the fact that the square of the Dirac 
Hamiltonian takes the form  $-\partial_x^2+\Delta_1^2+\sigma_3\Delta_1'$. The existing tools 
(see, e.g. \cite{Schubin}) for the analysis of the Schr\"odinger operators can be exploited to 
reveal spectral properties of the Dirac Hamiltonian. In this context, let us mention that 
interesting results were obtained by the spectral analysis of general class of deformed quantum 
waveguides described by Schr\"odinger equation \cite{quantumwaveguides}. We believe that 
similar analysis for the carbon nanostructures described by the one- or two-dimensional Dirac 
Hamiltonian would be fruitful. 

The presented analysis is qualitative. The equation (\ref{eq1}) is a good approximation for 
the quasiparticles in carbon nanotubes only for small region of the momentum space where the linear 
dispersion relation is valid. When the gap opened by the pseudo-magnetic field in the spectrum 
is too big, nonlinear (the so-called trigonal warping) terms \cite{trigonal} have to be included 
into the Hamiltonian.   
In the article, we neglected surface curvature of the nanotubes. The tubular surface prevents 
the $\pi$-orbitals of the carbon atoms to be parallel to each other. This implies presence of 
additional pseudo-magnetic fields in the Hamiltonian. However, in case of armchair nanotubes, 
these additional gauge fields can be transformed out \cite{KaneMele}.

The examples presented in the text suggest that the non-uniform radial twist can induce bound 
states in the nanotube. 
In particular, the second model with double-kink potential provides an interesting qualitative 
insight into realistic experimental setting: the nanotube with asymptotically vanishing twist 
is immersed into the homogeneous magnetic field. Besides the explicit formula (\ref{LDOSII}) 
for LDOS, the model predicts the appearance of bound states and the formula (\ref{induced}) 
fixes their energies in dependence on the asymptotic twist.  The model can be simply tuned 
with the use of perturbation techniques. 

The supersymmetry can be very useful for construction of the models with more complicated 
(yet asymptotically constant) twist inducing richer spectral properties.
The formalism presented in the second section can be repeated to produce a chain of solvable 
Hamiltonians, $h$, $\tilde{h}$, $\tilde{\tilde{h}},...$, by taking the last constructed operator 
as the seed Hamiltonian for the new system. These new solvable systems shall provide insight 
into the \textit{deformation-induced spectral engineering} of carbon nanotubes. 
The reflectionless models are particularly important in this context; they are analytically 
feasible and possess nontrivial (super)symmetry, analog of (\ref{A}) and (\ref{dressedhiddensusy}).   
The considered double-kink example suggests that the number of bound states could be in a simple 
relation to the vector potential of the Hamiltonian; the number of bound states might be 
proportional to the number of minima of the potential. Verification of this hypothesis goes 
beyond the scope of the current paper and should be discussed elsewhere.

{\bf Acknowledgements:}

The work  has been partially supported
by FONDECYT Grant No. 1095027, Chile,
and  by the GA\v CR Grant P203/11/P038, Czech Republic.
VJ  thanks the Department of Physics of the Universidad de
Santiago de Chile for hospitality. %VJ is a member of Doppler Institute which is founded by grant LS06002.


\begin{thebibliography}{99}
\bibitem{Paulimatrix} The Pauli matrices $\sigma_a$, $a=1,2,3$, are defined in a standard way. %as $\sigma_1=\left(\begin{array}{cc}0&1\\1&0\end{array}\right)$, $\sigma_2=\left(\begin{array}{cc}0&-i\\i&0\end{array}\right)$ and $\sigma_3=\left(\begin{array}{cc}1&0\\0&-1\end{array}\right)$. 
Identity matrix is denoted as $\mathbf{1}$.
\bibitem{feinbergformule} Similar formula was found in \cite{Feinberg} for the reflectionless models by the inverse scattering method.
\bibitem{scale} Equation (\ref{eq1}) is the approximation of the tight-binding model 
for energies that are close to the selected Dirac point (i.e. $E\sim 0$).  In the 
tight-binding model of the nanotubes  \cite{Roche}, \cite{nasKlein}, the mass 
(or the external field) parameter $m$ is in the fixed relation to the radius $r$ of 
the nanotube.  The quantized momenta form equidistant lines with mutual distance 
$r^{-1}$ in the first Brioullin zone. In case of the metallic nanotubes, one of the 
lines passes through the considered Dirac point when no external field is present. 
The external field associated with $\tilde{\Delta}_{MG}$ induces the spectral gap by 
shifting this line into the distance $m$ from the Dirac point. Hence, $m$ corresponds 
to a fixed fraction of the distance between the lines. Recovering the relevant constants, 
we get $m=\frac{\sqrt{3}\,a_l\,\gamma_0}{2r}\,\beta$ and $\Delta_1=\gamma_0 s_{xy}$, 
\cite{ando}. Here $\gamma_0$ is hopping parameter ($\gamma_0=2.9\,eV$), the lattice 
constant $a_l\simeq 2.46\,\mbox{\AA{}}$ and $\beta\in [0,1/2]$ is dependent on the external 
field. For $\beta=\frac{1}{3}$, the system is equivalent to a semi-conducting nanotube 
in absence of the external field \cite{Roche}.
\end{thebibliography}
\end{document}